\documentclass[]{spie}  
\pdfoutput=1
\usepackage[]{graphicx}

\newcommand{\apj}{Astrophysical Journal}
\newcommand{\apjs}{Astrophysical Journal Supplement}
\newcommand{\mnras}{Monthly Notices of the Royal Astronomical Society}
\newcommand{\pasp}{Publications of the Astronomical Society of the
  Pacific}
\newcommand{\aap}{Astronomy \& Astrophysics}
\newcommand{\aj}{Astronomical Journal}

\newcommand{\antares}{\textsc{antares}}
\newcommand{\amelie}{\textsc{amelie}}

\title{ANTARES: A Prototype Transient Broker System} 

\author{Abhijit~Saha,\supit{a}
 Thomas~Matheson,\supit{a}
  Richard~Snodgrass,\supit{b}
  John~Kececioglu,\supit{b}
 Gautham~Narayan,\supit{a}
 Robert~Seaman,\supit{a}
 Tim Jenness,\supit{c}
 Tim~Axelrod\supit{b}
\skiplinehalf
\supit{a}NOAO, 950 N Cherry Avenue, Tucson, AZ 85719 USA; \\
\supit{b}University of Arizona, Tucson, AZ 85721 USA; \\
\supit{c}Department of Astronomy, Cornell University, Ithaca, NY, 14853, USA
}

\authorinfo{Further author information: (Send correspondence to A. Saha)\\ A.Saha: E-mail: saha@noao.edu, Telephone: +1 520 318 8288}

\begin{document} 
  \maketitle 

\begin{abstract}
The Arizona-NOAO Temporal Analysis and Response to Events System (ANTARES) is a joint project of the National Optical Astronomy Observatory and the Department of Computer Science at the University of Arizona. The goal is to build the software infrastructure necessary to process and filter alerts produced by time-domain surveys, with the ultimate source of such alerts being the Large Synoptic Survey Telescope (LSST). The \antares\ broker will add value to alerts by annotating them with information from external sources such as previous surveys from across the electromagnetic spectrum. In addition, the temporal history of annotated alerts will provide further annotation for analysis. These alerts will go through a cascade of filters to select interesting candidates. For the prototype, `interesting' is defined as the rarest or most unusual alert, but future systems will accommodate multiple filtering goals. The system is designed to be flexible, allowing users to access the stream at multiple points throughout the process, and to insert custom filters where necessary. We describe the basic architecture of \antares\ and the principles that will guide development and implementation.
\end{abstract}




\section{The Problem}
\label{sec:intro}  

\subsection{Background}
 An increasing number of large area astronomy
surveys that probe time-variable phenomena are producing candidates
that need rapid response follow-up (whether monitoring ongoing
changes, or spectroscopic observations to probe physical
characteristics).  Salient examples are: the Lick
 Observatory
Supernova
Search\cite{2011MNRAS.412.1419L}\footnote{http://astro.berkeley.edu/bait/public\_html/kait.html},
the Catalina Real-Time Transient
Survey\cite{2009ApJ...696..870D}\footnote{http://crts.caltech.edu/}, the Panoramic Survey
Telescope \& Rapid Response
System (Pan-STARRS)\cite{2002SPIE.4836..154K}\footnote{http://pan-starrs.ifa.hawaii.edu/public/}, the
Palomar
 Transient Factory (PTF and
iPTF)\cite{2009PASP..121.1395L}\footnote{http://ptf.caltech.edu/iptf/}, and the La Silla-Quest
Variability Survey\cite{2012IAUS..285..324H}\footnote{http://hep.yale.edu/lasillaquest}.  These
surveys are discovering new transient phenomena that are already
taxing the available follow-up capacity of telescope facilities
world-wide.
 These projects have developed tools to filter their
discoveries to focus on
 events of interest to their research teams
(e.g., supernovae, gamma-ray burst events, and so on), which can gather
ancillary information about their `alerts' from external catalogs, and
use the available information to classify the sources associated with
their alerts. A leading example of this is
SkyAlert\cite{2009ASPC..411..115W}\footnote{http://skyalert.org/}, a system that has solved
many
 of the astronomical issues associated with adding value to
alerts. SkyAlert enables users to create filters on alerts, including
ancillary information on these alerts, in order to find relevant
events.  The PTF system also employs tools to identify interesting
alerts~\cite{bloom12}.  The scale of time-domain alert generation,
though, is quickly increasing.  The Zwicky Transient
Facility~\cite{kulkarni12} (ZTF) will have more than 6 times the
field-of-view
 of PTF, while time domain surveys with DECam on the
Blanco telescope
 benefit not only from the 3 deg$^2$ field-of-view,
but the depth
 attainable with a 4m-class facility.  Moreover,
transients are
 generated across the electromagnetic spectrum, from
radio facilities
 such as LOFAR\cite{2013A&A...556A...2V}\footnote{http://www.transientskp.org/}
to high-energy
 space-based observatories such as
Fermi\cite{2009ApJ...697.1071A}\footnote{http://fermi.gsfc.nasa.gov/}, making the overall
problem that much more complex.
 
 On the
horizon for beginning operation in 2021, is the Large Synoptic Survey
Telescope \cite{krabb10,2009arXiv0912.0201L}.
 With its ~10 deg$^2$ field-of-view and
$\sim$6m collecting area, the transient detection
 rate leaps by
orders of magnitude.  LSST will detect (with 5$\sigma$
 significance)
$10^3-10^4$ alerts per image, or $10^6-10^7$ per night.
 By going
fainter, and covering an area of over 18,000 square degrees, this 10
year long survey will probe an unprecedented volume of space with a
time cadence that can identify variability on time scales from tens of
minutes to years.  During the survey operation, the
LSST facility will issue alerts of celestial transient events using
VOEvent and other IVOA
protocols\cite{2006ASPC..351..637W,2011arXiv1110.0523S}.

\subsection{Variable Event Alerts}
 
 An alert is a notice triggered
when an image shows that something is significantly different with
respect to an archive image.  A variable star may trigger an alert
each and every time it is imaged: a supernova in a distant galaxy will
trigger repeatedly against an archive image from before it erupted.  A
moving object will be seen typically over erstwhile blank sky, and
move to a different location at subsequent 
 epochs, triggering alerts
at all these different locations at each respective epoch.
 While most
alerts will be yet another incremental data point for a celestial
object already known to vary, among these multitudes will lurk objects
the likes of which have never (or extremely rarely) been seen before.

 Prompted by the potential importance of early detection of short
lived transient phenomena, the LSST survey will issue alerts with a
latency of only about a minute. An alert contains essential
information like the location on the sky, the passband in which the
variability was detected, whether the change was in brightness or in
position, the magnitude of the change, and the epoch of that
particular trigger.  It may or may not (depending on the facility that
issues it) contain ancillary information about whether it is a
recurrent alert, a history of all alerts at that location in the sky, 
or other similar ancillary information.  
\begin{figure}[ht!]
\begin{center}
\includegraphics[height=20cm]{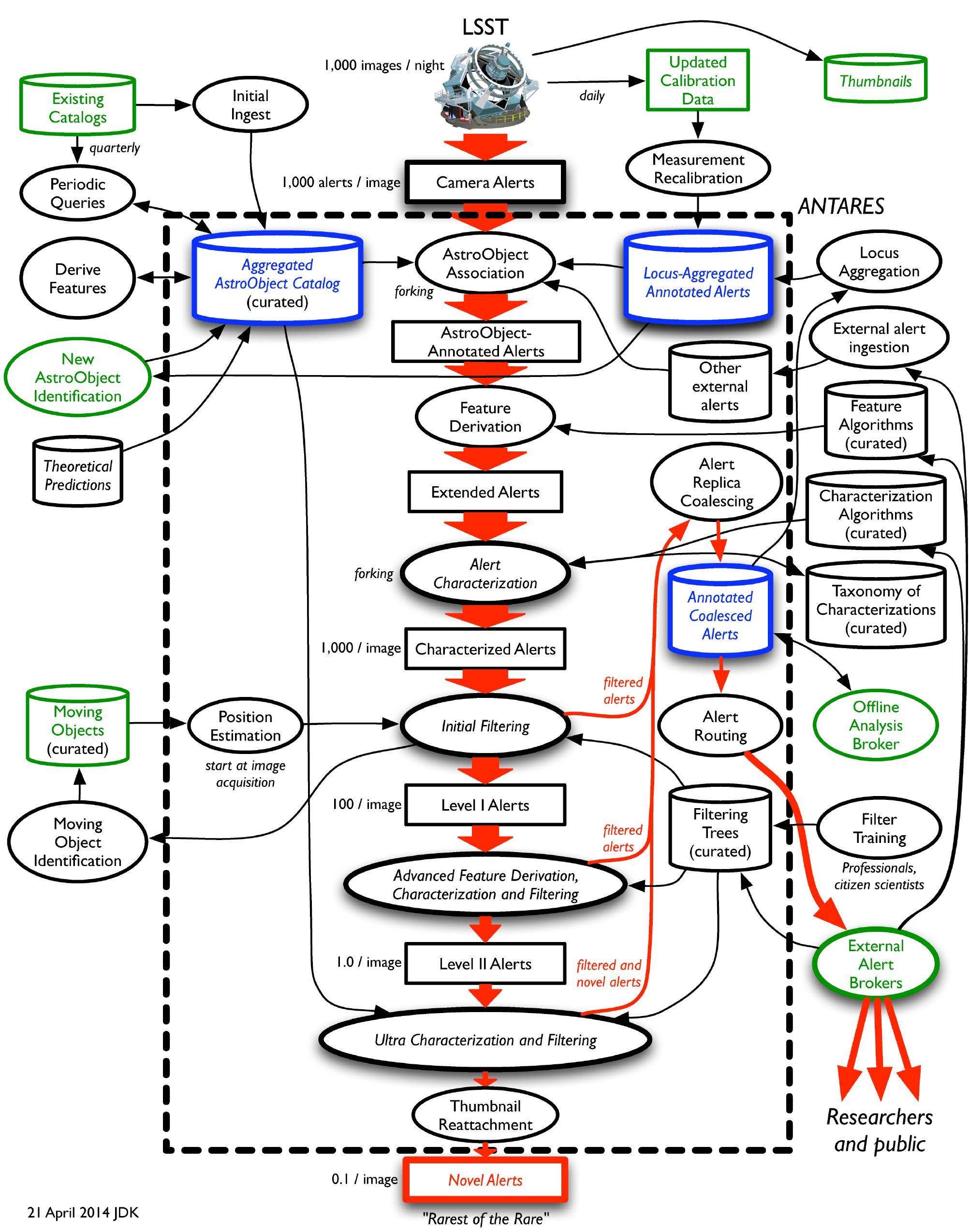}
\caption{Basic architecture of the \antares\ system.  The dashed box
  encompasses the processes that must keep up with the LSST
  frame-rate.}
\label{fig:antares}
\end{center}
\end{figure}

 A good fraction of alerts
from LSST will be known variable stars or moving
objects~\cite{ridgway12,grav11}, but hidden among them will be rare
and interesting
 objects that have relatively short lifetimes. Only
with additional
 follow-up will these objects reveal their nature.
These could range
 from short-lived phases of stellar evolution such
as the final helium
 flash~\cite{herwig05,iben83} to superluminous
supernovae~\cite{galyam12} to
 electromagnetic counterparts of LIGO
detections~\cite{sigg08,nissanke13}.
 Beyond these rare, but known or
predicted, objects lies the great
 discovery space that awaits LSST.
The superluminous supernovae were
 essentially unknown fifteen years
ago and the discovery of dark energy
 was certainly surprising.  Over
its life, LSST will generate more than
 a billion alerts and some will
be completely unknown and
 unanticipated objects.  Without the ability
to rapidly sort through
 millions of alerts each night and winnow them
down to a reasonable
 number that can be studied in detail, we will
lose these rare and
 potentially extraordinarily interesting objects.
The astronomical
 community is becoming more aware of the necessity of
such a
 tool~\cite{matheson13}. 

\subsection{Project Goals}
 
We take note of (and are encouraged by)
advanced methods for determining classification probabilities using
small numbers of time-series measurements (for instance \cite{richards11,
  angeloni14, masci14}), using machine learning algorithms and
techniques. They have been used to identify specific kinds of
variables of interest to their respective investigations. Their
successes notwithstanding, the alert rates that the LSST survey
promises require us to look at the problem a little differently:
\begin{itemize}

\item
The alert broker needs to handle alert data volumes at the rate that LSST is capable of generating, i.e. thousands of alerts per minute.  They need to be processed with a latency that does not make them stale:  i.e.  since LSST will produce alerts with a latency of $\sim 1$ minute, the broker must process those alerts without introducing significantly larger delays to the alert stream.
\item
The alert broker needs to serve generic needs: i.e. its design should not be limited to identifying specific pre-defined kinds of celestial sources. 
\item
It should store and archive all alerts, and be able to append contextual information for any celestial sourc(es) associated with that alert. It should be able to integrate results from any followup investigation of sources.

\end{itemize}

Achieving these goals requires the combined efforts of astronomers with experience in time-domain astronomy, as well as computer scientists who can impose the design and methods necessary  to achieve the necessary end-to-end speed and scalability for dealing with LSST scale data rates and volumes.

 \section{Design and Architecture of ANTARES}

\subsection{Proto-type design considerations}
 
The knowledge we have about an alert, such as brightness, change in
flux, Galactic coordinates, ecliptic coordinates, distance to nearest
galaxy, etc., constitute features that can probabilistically
characterize alerts.  We re-iterate that this is a broad
characterization, not a specific classification: the latter will
have to come from software systems further downstream.  Because of the
time-scale of LSST exposures, with a new image every $\sim$37 seconds,
alerts must be processed rapidly to keep up with the data stream.
Classification often requires more complex analysis and usually a more
complete light curve~\cite{richards11,graham13}.

For the prototype, we have selected the challenging problem of identifying the rarest of 
time domain phenomena: those that are least like things we know.  Alerts that appear to come 
from more commonplace astronomical sources are diverted,
but saved for further use.  We will discuss later how \antares\ is structured so that it can be modified and applied to identify other kinds of phenomena, and thus become a generic tool.  Identifying the `rarest of the rare' leads us through the problem space that makes adapting to other needs relatively straight-forward.

\subsection{Architecture and Data Path}

Figure~\ref{fig:antares}
illustrates the main components of the \antares\ architecture\cite{2014AAS...22334302M,HWTUIII_Matheson}.  The
overall design principles are open source and open access.  The
software will be available for anyone to implement and our
implementation will be community driven.  The alert stream can be
tapped at many points throughout the system. In
Figure~\ref{fig:antares} alerts enter the system from the top center.

The first stages provide  annotation that add contextual value to
the alerts.  Source association is a critical step to incorporate
relevant astronomical knowledge for each alert.  Catalogs of
astronomical information, as well as the LSST source catalog will be
the basis for this source association.  Examples include the 2MASS
All-Sky Data
Release\cite{2006AJ....131.1163S}\footnote{http://www.ipac.caltech.edu/2mass/releases/allsky/},
the Chandra Source
Catalog\cite{2010ApJS..189...37E}\footnote[1]{http://cxc.harvard.edu/csc/index.html}, the NRAO
VLA Sky Survey\cite{1998AJ....115.1693C}\footnote[2]{http://www.cv.nrao.edu/nvss/}, the Sloan
Digital Sky Survey\cite{2009ApJS..182..543A}\footnote[3]{http://www.sdss.org/}, the NASA
Extragalactic Database\cite{1991ASSL..171...89H}\footnote[4]{http://ned.ipac.caltech.edu/},
and
GAIA\cite{2009MmSAI..80...97C}\footnote[5]{http://sci.esa.int/science-e/www/area/index.cfm?fareaid=26},
among many others.  These external catalogs are collated and initially
ingested to produce the {\it Aggregated AstroObject Catalog}, shown 
near the top-left of Figure~\ref{fig:antares}.  This catalog will be
updated from time to time with periodic queries to the external
catalogs, and with new `AstroObjects' from the episodic  data
releases of the LSST survey.  Even the proximity to known sources can
provide useful constraints.

A new alert is also tested for
association with past alerts (from a database maintained by \antares ,
and shown in the figure as {\it Locus-Aggregated Annotated Alerts})
and additionally from any other available external alert data sources.
The history of flux measurements, such as a light curve, will be
valuable annotation.  The {\it Locus-Aggregated Annotated Alerts} is
meant to be an efficient database that can be updated regularly is an
essential element of the system.  This will be a valuable
astronomical resource on its own.  As mentioned before, the SkyAlert
system provides a similar annotation.  The problem for the future is
the scale of alerts and the resulting necessity of this efficient
database being integrated into the system for brokering alerts.

 A central notion in our procedure is that of {\it alert
  characterization}.  This is a discriminant activity  which uses the
features to determine what `kind' of alert we have.  We distinguish
this task from `classiﬁcation', in that characterization is
necessarily uncertain and probabilistic, while classiﬁcation is a
more certain association with a known astrophysical type.  Examples of
broad characterizations include known variable star, extragalactic
source, active galaxy, or likely moving object. These require looking
at all of the features, and as such is a holistic analysis, as
contrasted with feature derivation, which can be performed
independently for each added feature. For example, a small change in
magnitude might imply a stellar variable, but if it has not been
detected before, and it is near a galaxy, it may be a supernova, but
caught when the brightness is changing slowly.

For many alerts,
there will only be a small number of features available for
characterization, especially for an initial detection. If there are
not enough features for discrimination by filtering, we can apply a
probabilistic expectation of variability based on position on the sky
and known distributions of variability~\cite{ridgway12}.  For a
position, we can construct a variability probability density function
and predict the likelihood of the alert as observed.  

With more
data, more features become available and more complex filtering
algorithms can be used. \antares\ will then use multiple layers of
filters to sort the alerts and find the rarest or most interesting
among them (the focus of the prototype project).  The filtering will
be based on feature vectors, either directly supplied  by the alert
and associated contextual information, or derived therefrom. These
features  are then compared against the features from known
time-variable phenomena, using a variety  of methods. Alerts that are
likely to come from `common-place' phenomena are diverted away from
the main processing stream.  Each stage lets through fewer and
progressively less commonly characterized  alerts.  These may then be
re-characterized in feature vector space that is different or of
higher dimension and filtered again.   The filtering stages are meant
to be ordered so that most efficient (decisions on most alerts in the
least time) filters are staged first. More time consuming and in-depth
probing is reserved for the later stages, where the alerts have
already been winnowed to a smaller number.  Experimentation will show
us the most efficacious and efficient feature combinations and
algorithms.   The training of filters and algorithms will be aided
using machine based experimentation with \amelie\ (see below in
section~\ref{subsec:amelie}), and  is an integral part of the
development of the \antares\ system.

The {\bf diverted alerts are
  not discarded:} in each filtering stage they are diverted from the
main filtering stream but are still accessible to other filtering
systems, including, potentially, copies of the \antares\ system
itself that are tuned to other specific goals.  Thus an {\it External
  Alert Broker}  (shown in the bottom right of
Figure~\ref{fig:antares}) can utilize the value added material from
\antares\ to filter according to alternative needs and priorities.
Custom filters can be applied, allowing users to isolate exactly
which of the alerts is of interest to them and thus address many
different goals.  These community-derived filtering algorithms will
be applied in a multi-step process, allowing for better management of
computational resources.  By characterizing the alerts, the number of
dimensions of feature space can be reduced.  More complex filters can
be applied to the smaller number of alerts after initial filtering
stages.

An important design consideration throughout the
architecture of \antares\  is the structured provision of community
input. While \antares\  will provide the overarching design of the alert
analysis, it is the role of the astrophysical community to
provide the specific algorithms used at various places along the
filtering process.
\begin{figure}[h]
\begin{center}
\includegraphics[height=9cm]{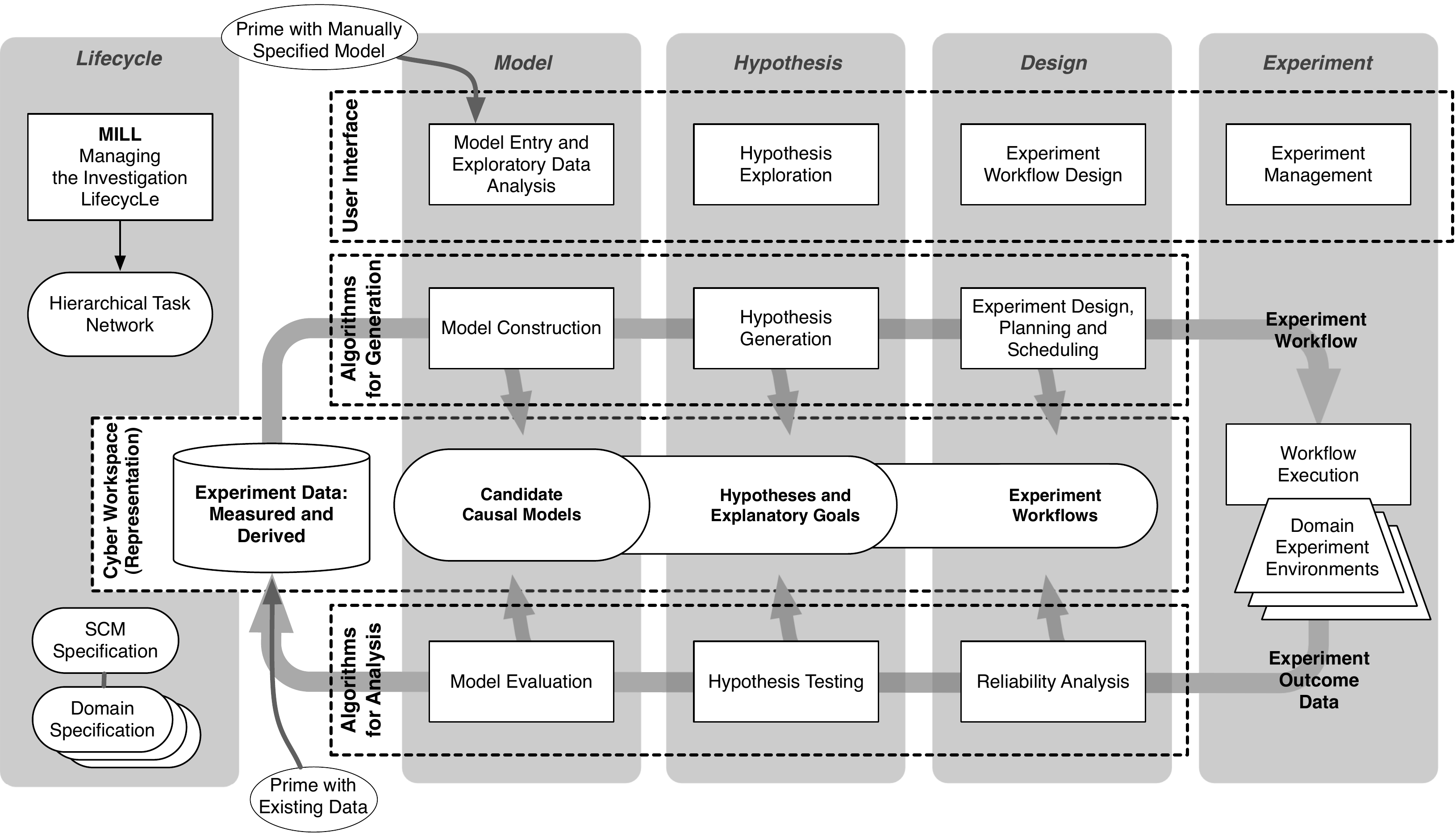}
\caption{Basic architecture of \amelie.}
\label{fig:amelie}
\end{center}
\end{figure}

\subsection{AMELIE}
\label{subsec:amelie}

The Arizona Machine-Experimentation Laboratory (\amelie), Figure~\ref{fig:amelie}), provides a system for
constructing and testing
structural-causal models~\cite{morrison11}.  This essentially automates
the scientific process and allows us to run experiments to test
relationships among features, including relationships that have not
yet been apparent.  It can observe the operation of \antares\
and make it more efficient.

\subsection{From Proto-type to a Generic Tuneable Broker}

The goal for the prototype is to distinguish rare and unusual objects.
Once it is operational, the next stage is to expand the scope to allow
users to find any type of alert of interest to them.  We foresee that 
there will be many stages of the \antares\ system itself, processing
different data streams over different time scales.  The overall alert
ecosystem could accommodate multiple alert input streams and thus find
a general way to serve the astronomical community's needs.

\acknowledgements 
 We acknowledge the NSF INSPIRE grant (CISE AST-1344204, PI:Snodgrass)
that supports this work.  AS acknowledges many conversations with Kirk Borne,
which seeded much of the thinking from which this project was born.

\end{document}